\documentclass[aps,prl,twocolumn,showpacs,superscriptaddress]{revtex4-1}

\usepackage{graphics}
\usepackage{graphicx}
\usepackage{dcolumn}
\usepackage{bm}
\usepackage{color}
\usepackage{amssymb,amsmath}

\begin{document}

\title{Soliton-radiation trapping in gas-filled photonic crystal fibers}
\author{Mohammed F. Saleh}
\affiliation{Max Planck Institute for the Science of Light, G\"{u}nther-Scharowsky str. 1, 91058 Erlangen, Germany}
\author{Fabio Biancalana}
\affiliation{Max Planck Institute for the Science of Light, G\"{u}nther-Scharowsky str. 1, 91058 Erlangen, Germany}
\affiliation{School of Engineering and Physical Sciences, Heriot-Watt University, EH14 4AS Edinburgh, UK}
\date{\today}

\begin{abstract}
We propose an optical trapping technique in which a fundamental soliton traps an ultrashort small amplitude radiation in a symmetric hollow-core photonic crystal fiber filled with a noble gas, preventing its dispersion. The system is Raman- and plasma-free. Trapping is due to the cross phase modulation effect between the two pulses. The trapped radiation inside the soliton-induced potential will oscillate periodically due to the shock effect, similar to the motion of a mechanical pendulum.
\end{abstract}
\maketitle

\paragraph{Introduction ---}
Optical trapping has been observed in a variety of different situations since the invention of low-loss optical fibers \cite{French74}. Self-trapping is at the core of the existence of optical solitons, where the nonlinear self-phase modulation (SPM) is used to compensate the group velocity dispersion (GVD) \cite{Agrawal07}. Bright and dark solitons have been observed in the anomalous and the normal dispersion regimes, respectively, in a number of elegant experiments \cite{Mollenauer80,Mollenauer83,Weiner88}. Trapping can also take place between a pair of pulses, when they lie in opposite dispersion regimes to satisfy the group velocity matching condition. For instance, red-shifted solitons due to intrapulse Raman can trap across the zero dispersion wavelength weak blue-shifted dispersive wave-packets on the femtosecond scale \cite{Nishizawa02a,Nishizawa02b,Hori04,Genty04,Gorbach06,Gorbach07a,Gorbach07b}. In an unconventional situation, a bright soliton can also be trapped in the normal dispersion, when it is coupled to a dark soliton in the anomalous dispersion regime via the nonlinear cross-phase modulation (XPM) effect \cite{Trillo88,Kivshar92}.  

The invention of photonic crystal fibers (PCFs) is a milestone in the field of nonlinear fiber optics \cite{Russell03,Russell06}. Hollow-core (HC) PCFs have been exploited to explore the nonlinear interaction of light in a new wide-range of distinct media. HC-PCFs based on a Kagome lattice have been successfully used recently in the investigation of light-matter interactions in presence of gases \cite{Travers11}. Filling these fibers by a Raman-active gas, a drastically reduction in Raman threshold has been obtained \cite{Benabid02a}. High harmonic, and efficient deep UV generation have been also demonstrated in HC-PCFs filled by noble (Raman-inactive) gases \cite{Heckl09,Joly11}. Using intense broadband pulses to access the ionization regime of the noble gases, soliton self-frequency blueshift has been observed and thoroughly studied in these fibers \cite{Hoelzer11b,Chang11,Saleh11a,Saleh11b}. On the other hand, launching intense narrowband pulses can lead to ionization-induced asymmetric SPM and universal modulational instability \cite{Saleh12}.

In this Letter, we show that optical trapping between two different short pulses can exist, when \textit{both} pulses propagate together in the anomalous dispersion regime of a HC-PCF filled by a noble gas. The intensities of the two pulses are below the ionization limit of the gas.

\paragraph{Principle of operation ---}
Consider the propagation of two short pulses in a guided nonlinear-Kerr medium. Assuming that only one of the two pulses is intense enough to induce nonlinear phenomena, and the dispersion is anomalous for both pulses. Hence, the strong pulse can maintain its temporal shape due to the induced SPM, while the other pulse will suffer from strong dispersion due to the absence of compensation between nonlinear and dispersive effects. Only if the two pulses can propagate with the same group velocity, the nonlinearity of the strong pulse will affect the second pulse via XPM that can compensate the dispersion-induced broadening, and the weak pulse will propagate as a temporally localized radiation. In analogy with quantum mechanics, the strong pulse creates a potential well that traps the weak pulse inside it, and both pulses travel together without suffering from dispersion-induced broadening \cite{Gorbach07a,Gorbach07b}. If the nonlinear medium is Raman- or plasma-free, the dispersion must be anomalous for both pulses to allow dispersion-compensation to take place.

\paragraph{System design ---}
To observe this temporal trapping due to the soliton-induced potential, we require a nonlinear-Kerr medium, where two different ultrashort pulses lying in the \textit{same} anomalous dispersion regime can acquire a common group velocity. In previous works, the group velocity matching condition was achieved when a red-shifting soliton was in the anomalous dispersion and the small amplitude wave was in the normal dispersion regime \cite{Gorbach07a,Gorbach07b}. One possible technique for our proposal is to use a solid-core PCF with multiple zero dispersion wavelengths. However, this type of fibers requires high fabrication tolerance. Moreover, undesirable emission of resonant radiation will be out of control in these fibers. Another technique is to launch the two pulses in two different polarization states in an elliptical solid core fibers. However, relatively long pulses must be used in this case to minimize the effect of Raman -- and it is that case one lacks the attracting feature of trapping ultrashort pulses.

A HC-PCF filled by a noble (Raman-inactive) gas and with a symmetric core is considered to be the perfect host to demonstrate our proposal. This can be achieved by launching two different pulses with the same frequency, however in two different circular polarization states, in the deep anomalous dispersion regime of the fiber. Having the same frequency allows the two pulses to have the same group velocity, as well as to be naturally in the same dispersion regime. Using different polarization states allows to monitor the two pulses at the fiber output. The reason for using circular polarization states is to avoid unwanted additional nonlinear phenomena to emerge if linearly polarized states are injected \cite{Agrawal07}. 

\paragraph{Governing equations for soliton-radiation trapping ---}
To study the induced trapping between a propagating strong pulse and a small amplitude radiation with the same central frequency and different circular polarization states in a lossless nonlinear medium, the following set of coupled nonlinear Schr\"{o}dinger equations can be used \cite{Agrawal07},
\begin{equation}
\begin{array}{l}
\left[ i\partial_{z}+\hat{D}_{1}(i\partial_{t})+\frac{2}{3}\gamma\left( \left|A_{1}\right|^{2}+ 2 \left|A_{2}\right|^{2}\right)\right]A_{1} = \frac{i}{2}\Delta\beta A_{2}, \\ 
\left[ i\partial_{z}+\hat{D}_{2}(i\partial_{t})+\frac{2}{3}\gamma\left( \left|A_{2}\right|^{2}+ 2 \left|A_{1}\right|^{2}\right)\right]A_{2} = \frac{i}{2} \Delta\beta A_{1},
\end{array} \label{eq1}
\end{equation}
where $z$ is the longitudinal coordinate along the fiber, $ t $ is the time coordinate, $A_{j}\left( z,t\right)$ is the slowly varying amplitude of the $ j $th pulse, $ \hat{D}_{j}(i\partial_{t}) =i\beta_{1j}\partial_{t}-\frac{1}{2}\beta_{2j}\partial_{t}^{2}$,  $ \beta_{1j} $ and $ \beta_{2j} $ are the first- and second-order dispersion coefficients of the $ j $th pulse ($j=1,2$), higher order dispersion coefficients are neglected, $ \gamma $ is the nonlinear Kerr coefficient, and $ \Delta\beta $ is the fiber birefringence. $ A_{1} $ and $ A_{2} $ correspond to the strong pulse and the weak radiation, respectively. In our proposed system, the medium is Raman- and plasma-free, $ \beta_{11} =\beta_{12}  $, $ \beta_{21} =\beta_{22}  $, and $ \Delta\beta =0 $ for perfectly symmetric cores. In each equation, the first (second) nonlinear term represents the self- (cross-) phase modulation effect. We introduce the following rescalings: $ \xi=z/z_{0} $, $ \tau=t/t_{0} $, $ \psi_{1}=A_{1}/A_{0} $, $ \psi_{2}=A_{2}/A_{0} $, $ A_{0}^{2}=3/(2\,\gamma z_{0})$, $ z_{0}=t_{0}^{2}/\left|\beta_{21}\right| $, where $ t_{0} $ is the input pulse duration. In  a reference frame moving with a group velocity $ v_{g} $, which is defined as $ v_{g}=1/\beta_{11} $, the two coupled equations can be replaced by
\begin{equation}
\begin{array}{l}
i\partial_{\xi}\psi_{1}+\frac{1}{2}\partial_{\tau}^{2}\psi_{1}+ \left|\psi_{1}\right|^{2} \psi_{1} = 0 , \\ 
 i\partial_{\xi}\psi_{2}+\frac{1}{2}\partial_{\tau}^{2}\psi_{2}+2 \left|\psi_{1}\right|^{2}\psi_{2} = 0, 
\end{array} \label{eq2}
\end{equation}
where the nonlinearity of the weak radiation is neglected. The solution is well-known for the first equation as a nonlinear Schr\"{o}dinger soliton, $ \psi_{1}\left( \xi,\tau\right)=\eta\, \mathrm{sech}\left( \eta \tau\right)  \mathrm{exp}\,\left(i \eta^{2}\xi /2 \right) $ , with amplitude $ \eta $ \cite{Agrawal07}. We assume that the solution of the second equation has the form $\psi_{2}(\xi,\tau)=\psi_{0}\, f\left(\tau \right)  \mathrm{exp}\,\left(i q \xi \right) $, where $ \psi_{0} $ and $ q $ are the amplitude and the propagation constant of a pulse with temporal profile $ f\left(\tau \right) $. Substituting this form in the second equation, we obtain a linear one-dimensional Schr\"{o}dinger problem in time,
\begin{equation}
-\frac{1}{2}\:\partial_{\tau}^{2}f+U \left(\tau\right) f=-q\,f,\label{eq3}
\end{equation}
where $ U \left(\tau\right)=-2 \:\eta ^{2}\mathrm{sech}^{2}\left( \eta\, \tau\right)$ represents the potential well, $ f\left(\tau \right) $ is the eigenfunction, and $ -q $ is the corresponding eigenvalue that represents a discrete energy level. We have found the analytical expressions of the modes of this potential well. The fundamental (even) mode is $ f_{0}\left( \tau\right)=\mathrm{sech}^{p_{0}}\left( a \tau\right) $ with $ q= \frac{1}{2} \,\eta^{2}p^{2}_{0}  $, and $ p_{0}=\frac{1}{2}\left[ -1+\sqrt{17}\right]  $. The first-order (odd) mode behaves as $ f_{1}\left( \tau\right)=\mathrm{sech}^{p_{1}}\left( a \tau\right) \mathrm{tanh}\left( a \tau\right)$, with $q= \frac{1}{2} \,\eta^{2}\,\left(2-3 p_{1} \right)$, and $ p_{1}=p_{0}-1 $. For this system, \textit{there are only two localized modes}. In principle, $ \psi_{0} $ for either the fundamental or the first-order mode can take an arbitrary value as long as $ \psi_{0} \ll \eta$, since these modes propagate linearly. The problem can also be solved numerically by using sparse matrices \cite{Benham01}. The obtained numerical results validate the analytical formulas for the eigenfunctions and eigenvalues given above. It is worth mentioning that an attempted solution of a similar problem is presented in an eminent quantum mechanics textbook \cite{Landau77}, however, a full analytical form of the eigenfunctions is presented for the first time in this work.

\begin{figure}
\includegraphics[width=8.6cm]{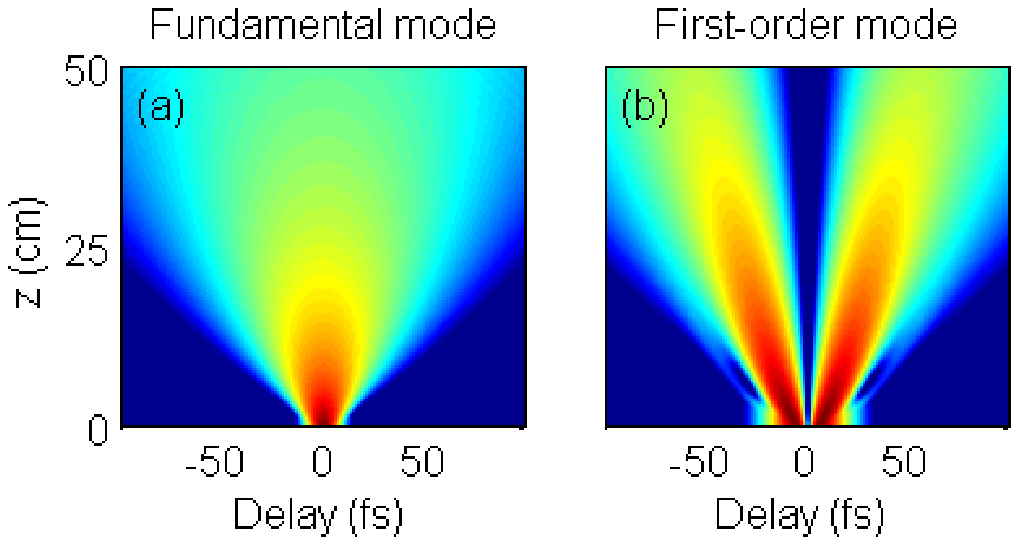}\vspace{-3.4cm}
\includegraphics[width=8.6cm]{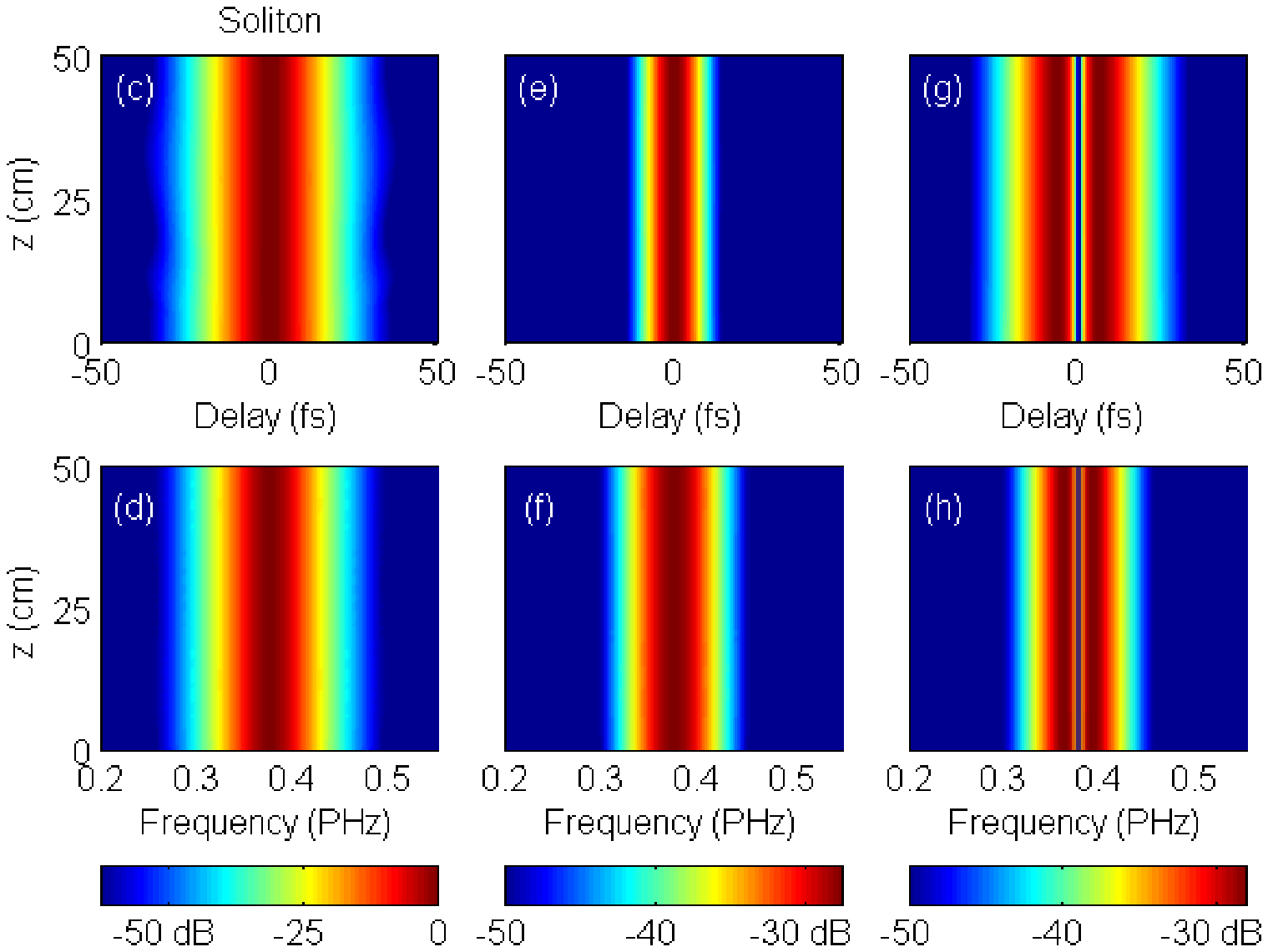}
\caption{(Color online). (a,b) Propagation of a dispersive radiation solely in an argon-filled HC-PCF, when the radiation takes the form of (a) the fundamental mode $ f_{0}\left(\tau \right)$, or (b) the first-order-mode $ f_{1}\left( \tau\right)$. (c-h) Propagation of a fundamental soliton and a dispersive radiation together in an argon-filled HC-PCF. The group-velocity matching condition is satisfied. The two pulses have the same central wavelength $\lambda=800$ nm, and opposite circular polarizations. (c,d) Temporal and spectral evolution of the soliton. (e-h) Temporal and spectral evolution of the radiation, when it is in the fundamental mode (e,f), or in the first-order-mode (g,h). Other simulation parameters are $ \eta=8.9 $,  $ \psi_{0} = 0.5$, and $ t_{0} = 50$ fs. Higher-order dispersion coefficients and self-steepening are neglected.
\label{Fig1}}
\end{figure}

Propagation of a dispersive radiation in an argon-filled HC-PCF is depicted in Fig. \ref{Fig1}. Panels (a,b) show the dispersion-induced broadening of the radiation in absence of the soliton-induced potential, when the launched pulse takes the shape of either the fundamental or the first-order mode, respectively. In presence of a fundamental soliton [with a temporal and spectral evolution shown in panels (c,d)], an XPM-induced trapping for the radiation can take place. Panels (e,f) and (g,h) present the propagation of a localized radiation, when it takes the form of the fundamental or the first-order mode, respectively. Higher-order dispersion coefficients and self-steeping effects are neglected in this simulation.

\paragraph{Shock effect ---}
Since the nonlinear medium is Raman and plasma free, and the two pulses are supposed to be arbitrarily short, the self-steeping (shock effect) will have a significant role in the pulse dynamics. In this case, Eq. (\ref{eq2}) becomes
\begin{equation}
\begin{array}{l}
i\partial_{\xi}\psi_{1}+\frac{1}{2}\partial_{\tau}^{2}\psi_{1}+\left|\psi_{1}\right|^{2}\psi_{1}=-i\, \tau_{\rm sh}\,\partial_{\tau}\left( \left|\psi_{1}\right|^{2}\psi_{1}\right) ,\\
i\partial_{\xi}\psi_{2}+\frac{1}{2}\:\partial_{\tau}^{2}\psi_{2}+2\left|\psi_{1}\right|^{2}\psi_{2}=-i\, 2\, \tau_{\rm sh}\,\partial_{\tau}\left( \left|\psi_{1}\right|^{2}\psi_{2}\right) ,
\end{array}\label{eq4}
\end{equation}
where $ \tau_{\rm sh}\equiv(\omega_{0}t_{0})^{-1}$ is the normalized shock coefficient, and $\omega_{0}$ is the input pulse central frequency. In order to study the shock effect on both the soliton and the radiation, we will adapt the variational perturbation method for our problem \cite{Agrawal07}. Using this method, the RHS of Eqs. (\ref{eq4}) are considered as perturbations for the previously obtained solutions of $ \psi_{1} $ and $ \psi_{2} $. Hence, the soliton and the radiation will maintain their functional shape during propagation, however, their parameters will vary. For both pulses, one has to obtain the Lagrangian $ \mathcal{L}=\int_{-\infty}^{\infty}\mathcal{L}_{d}\,d\tau $, where $ \mathcal{L}_{d} $ is the Lagrangian density. Using the reduced Euler-Lagrange equations, a set of ordinary differential equations that tracks the spatial evolution of the pulse parameters can be obtained. Since this technique is well illustrated in the literature for the soliton case \cite{Agrawal07}, we will show only its final result, $ \psi_{1}\left( \xi,\tau\right)=\eta\, \mathrm{sech}\left[ \eta\left(  \tau-\beta'\xi\right) \right]  \mathrm{exp}\,\left(i \eta^{2}\xi /2 \right) $, with $ \beta'=\tau_{\rm sh}\,\eta^{2} $. This means that the soliton acquires only a linear temporal shift during propagation due to the shock term. For the radiation,
\begin{equation}
\mathcal{L}_{d}=-\mathrm{Im}\left\lbrace \psi_{2}\,\partial_{\xi}\psi_{2}^{*}+2\,\varepsilon\,\psi_{2}^{*} \right\rbrace+\dfrac{ 1}{2}\left|\partial_{\tau}\psi_{1}\right|^{2}-2\left|\psi_{1}\right|^{2}\left|\psi_{2}\right|^{2} ,
\end{equation}
where $ \varepsilon= -2\, \tau_{\rm sh}\,\partial_{\tau}\left( \left|\psi_{1}\right|^{2}\psi_{2}\right)$,  $\mathrm{Im} $ stands for the imaginary part. Assuming that the radiation is in the fundamental mode,
\begin{equation}
\psi_{2}\left( \xi,\tau\right)=\psi_{0}\, \mathrm{sech}^{p_{0}}\left[ a\left(  \tau-\tau_{p}\right)\right]   \mathrm{exp}\,\left[-i\delta\left(\tau-\tau_{p} \right) \right],
\end{equation}
where $ \tau_{p} $ and $ \delta $ are the shifts due to the applied perturbation in the temporal and spectral position of the pulse center, respectively. After calculating the Lagrangian and applying the reduced Euler-Lagrange equations, one can show that the following equations describe the variation of the parameters:
\begin{equation}
\begin{array}{rl}
\dfrac{d\delta}{d\xi}= &- \dfrac{2\,\eta^{2}}{\psi_{0}^{2}\,I}\displaystyle\int_{-\infty}^{\infty}\!\!\!\! g\left( \tau\right)\left( \mathrm{Im}\left\lbrace \varepsilon\,\psi_{2}^{*}\right\rbrace+2\left|\psi_{1}\right|^{2}\left|\psi_{2}\right|^{2}\right)   d\tau ,\vspace{2mm} \\ 
 \dfrac{d\tau_{p}}{d\xi}=& -\delta+\dfrac{2\,\eta}{\psi_{0}^{2}\,I}\displaystyle\int_{-\infty}^{\infty} \,\mathrm{Re}\left\lbrace \varepsilon\,\psi_{2}^{*}\left(  \tau-\tau_{p}\right)\right\rbrace d\tau ,
\end{array} 
\label{equations7}
\end{equation}
where $ I\equiv\sqrt{\pi}\,\Gamma\left(p_{0} \right) /\,\Gamma\left(p_{0}+\frac{1}{2} \right) $, $ g\left( \tau\right) \equiv p_{0}\, \mathrm{tanh}\left[ a\left(  \tau-\tau_{p}\right)\right]$,   $ \Gamma $ is the gamma function, and $\mathrm{Re} $ stands for the real part. These coupled equations can be solved numerically. 

\begin{figure}
\includegraphics[width=8.6cm]{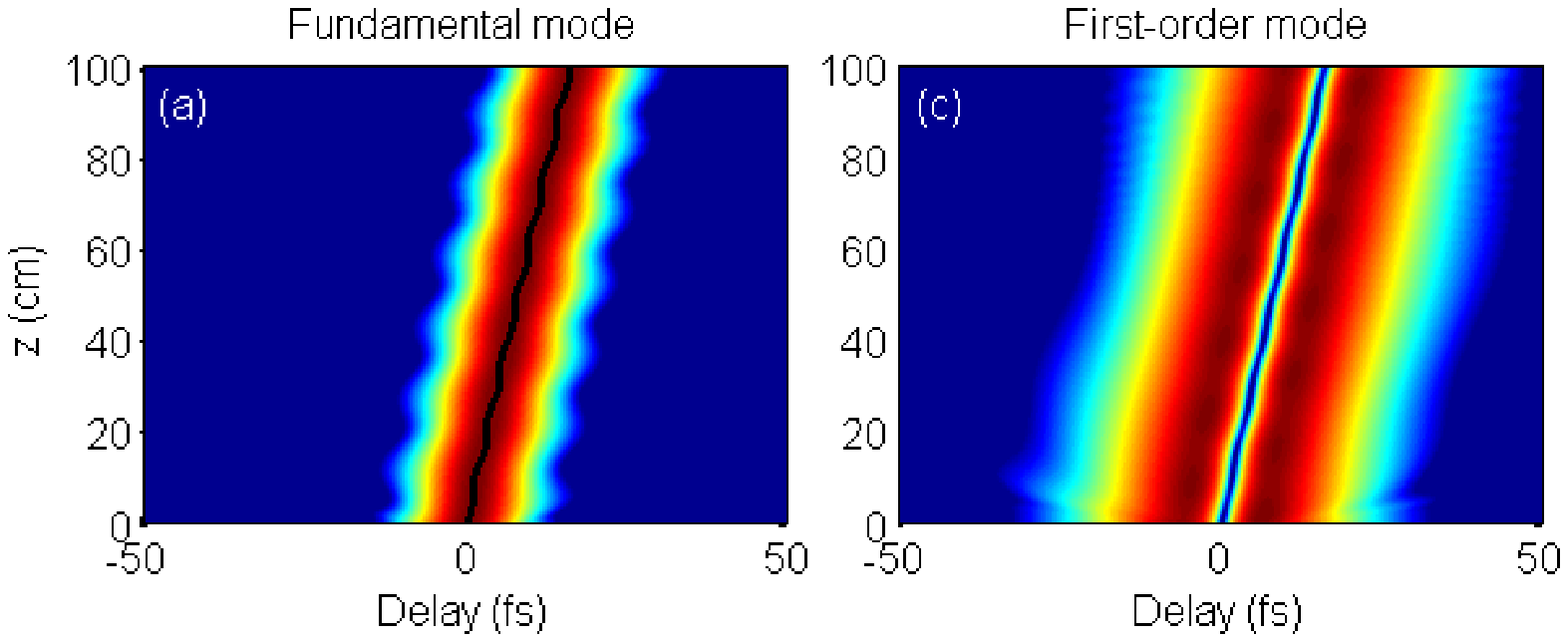}\vspace{-2.6cm}
\includegraphics[width=8.6cm]{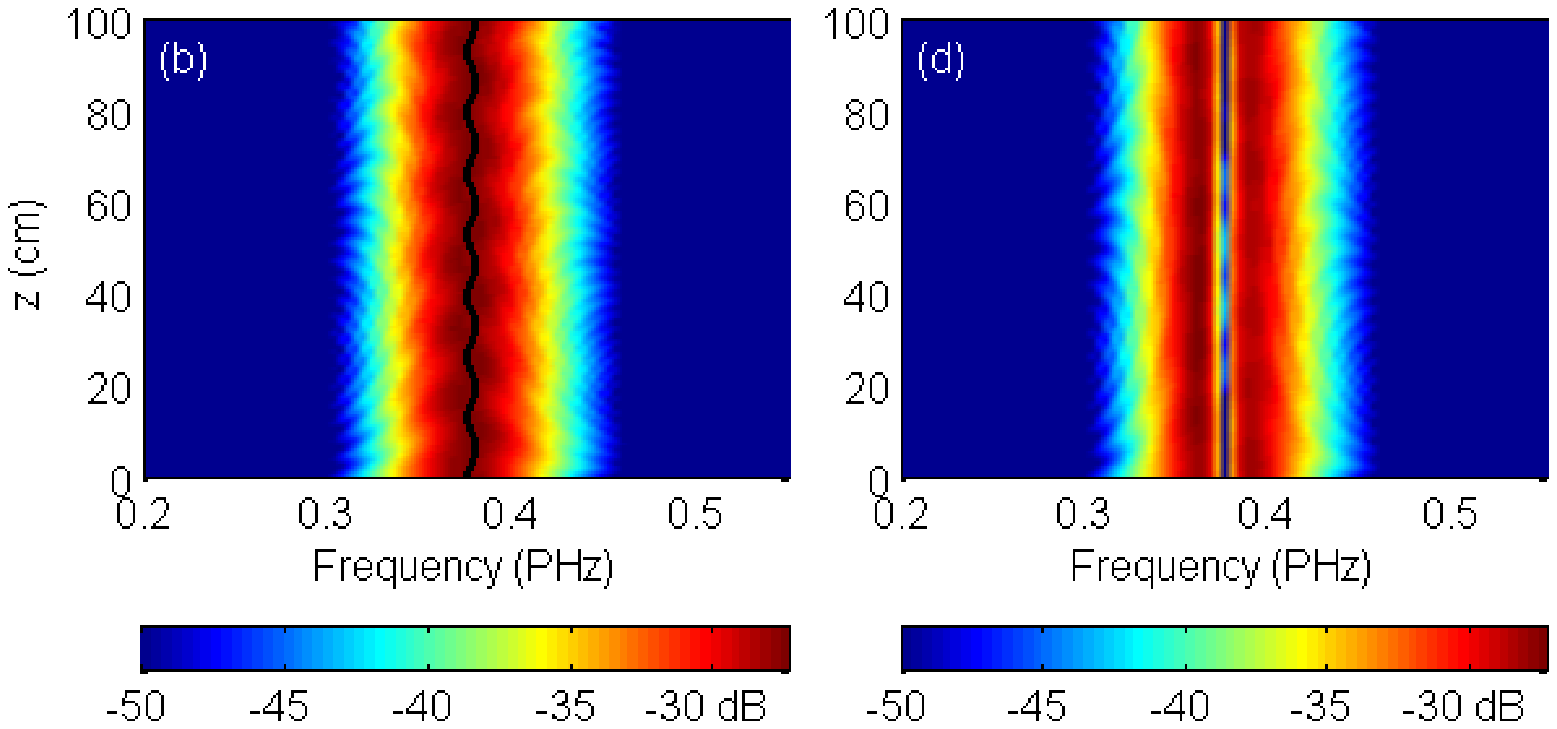}\vspace{-2cm}
\caption{(Color online). Propagation of a fundamental soliton and a dispersive radiation in an argon-filled HC-PCF in presence of self-steepening effect. Other simulation parameters are similar to those used in Fig. \ref{Fig1}. Temporal and spectral evolution of the radiation, when it is in the fundamental mode (a,b), or in the first-order mode (c,d), respectively. Solid black lines are the numerical solutions of the coupled differential equations Eq. (\ref{equations7}).
\label{Fig2}}
\end{figure}

The effect of the shock term on the radiation is depicted in Fig. \ref{Fig2}, which shows the temporal and the spectral evolution of the dispersive radiation. As shown in panel (a), the radiation acquires a varying group velocity, which differs from that of the soliton. The reason is related to the factor $2$ that distinguishes between the SPM and XPM processes [see Eq. (\ref{eq4})]. This group velocity difference between the two pulses causes the observed periodic oscillations of the radiation inside the soliton. The dynamics of the radiation is similar to the motion of a mechanical pendulum, or an oscillating particle inside a potential. Initially, the particle is settled at the potential minimum. Due to the shock effect, it acquires a velocity that is highest in correspondence with that minimum. As the `particle' (which is in our case the small amplitude radiation) moves upward, it undergoes a deceleration, resulting in a spectral blueshift as shown in panel (b). When the particle reaches its maximum position, it stops and changes its direction downwards, acquiring an acceleration that corresponds to a spectral redshift. Passing by the potential minimum, the particle goes upward to the other side of the potential. This periodic motion continues during the propagation along the fiber. In the case of the fundamental mode, there is a good qualitative agreement between the analytical and numerical results, see Fig. \ref{Fig2}(a,b). The odd mode behaves similarly to the even mode as portrayed in panels (c,d). However, the analytical results fail to mimic the numerical ones in this case. We attribute this to the reason that the radiation does not preserve its temporal shape during propagation. The two-lobes that constitute the radiation packet exchange energy, similar to partial coupling between adjacent waveguides, and this prevents the perturbation theory to be a good approximation to the dynamics.

\paragraph{Conclusions ---}
We present a trapping technique between a strong fundamental soliton and a weak dispersive radiation in a symmetric HC-PCF filled by a noble gas. The medium is Raman- and plasma-free. Trapping is due to the cross-phase modulation effect between the two pulses having the same central frequency but opposite circular polarization states.  Although both pulses lie in the deep anomalous dispersion regime of the fiber, this configuration satisfies unconventionally the group velocity matching condition. From quantum mechanics point-of-view, the soliton creates a potential well that traps the particle-like radiation inside, and the two pulses travel together without suffering dispersion-induced broadening.  This potential well allows only for two bound states. Exact analytical formulas for the fundamental and the first-order modes of the radiation have been obtained. The soliton shock effect leads to a periodic oscillation of the radiation inside the potential, similar to a pendulum motion. As a consequence, the radiation central frequency is shifted periodically.

\paragraph{Acknowledgements ---}
We would like to thank John C. Travers for useful discussions. This research is funded by the German Max Planck Society for the Advancement of Science (MPG).

\bibliographystyle{apsrev4-1}	

%

\end{document}